\documentclass[11pt, a4paper]{article}
\usepackage{amsmath, amsthm,  amsfonts, amscd}
\newcommand{\id}{{\text id}}
\renewcommand{\a}{\alpha}
\renewcommand{\b}{\beta}
\renewcommand{\c}{\gamma}
\renewcommand{\d}{\delta}

\newcommand{\isomorphism}{\cong }
\newcommand{\integers}{\mathbb Z }

\newcommand{\homotopyequivalence}{\simeq }
\newcommand{\inclusion}{\hookrightarrow }

\newcommand{\composition}{\circ}
\newcommand{\longto}{\longrightarrow}

\newcommand{\Z}{\mathbb Z}
\newcommand{\R}{\mathbb R}

\newcommand{\W}{{\cal W}}
\newcommand{\Ur}{{\cal U}_{res}}
\newcommand{\PU}{{\cal {PU}}}
\newcommand{\U}{{\mathcal{U}}({\mathcal{T}})}
\numberwithin{equation}{section}

\theoremstyle{plain}
\newtheorem{theorem}{Theorem}[section]
\newtheorem{lemma}{Lemma}[section]
\newtheorem{proposition}{Proposition}[section]

\theoremstyle{definition}

\newtheorem{note}{Note}[section]

\title{Principal Bundles and the Dixmier Douady Class}
\author{Alan L. Carey$^1$, Diarmuid Crowley$^2$, Michael K. Murray$^1$}
\date{4 March 1997 (Revised version)}
\begin{document}
\maketitle

\begin{abstract} A systematic consideration of the problem of
the reduction and extension of the structure group
of a principal bundle is made
 and a variety of techniques in each case are
explored and related to one another. We apply these to
the study of the Dixmier-Douady class in various contexts including
string structures, $\Ur$ bundles and other examples
motivated by considerations from quantum field theory.
\end{abstract}

\section{Introduction}
This paper develops
the theory of principal bundles with the aim of
studying various manifestations of the Dixmier-Douady class. The motivating
example is that of principal bundles
whose structure
group is an infinite dimensional Lie group much studied by many
authors
 in connection with string theory (it is the restricted
 unitary group in the terminology of Pressley and Segal \cite{PreSeg}).
Our results include a demonstration that there are
interesting examples of such bundles and we
relate them to string structures. We also
 discuss obstructions (or characteristic classes) arising from them.
 Finally we connect our work to the notion of bundle gerbe,
bundles with structure group the projective unitaries
and to
infinite dimensional Clifford bundles.
A statement of the main results of the paper
is given later in this introduction.

We now digress a little to explain the history which lead
to this paper.
Some time ago Gross \cite{Gro}
 suggested that quantum electrodynamics lends itself
to a formulation in terms of infinite dimensional Clifford bundles. It
was Segal \cite{Seg} who showed that using bundles with fibre the
projective space of a Fock space (carrying a representation of the
Clifford algebra) in non-abelian gauge
theories one could explain the origin of Hamiltonian anomalies
(in particular that discovered by Faddeev and Mickelsson \cite{FadMic}).
Mickelsson, in his study of anomalies and gauge theories,
found it useful to introduce the idea of a Fock bundle.
These are also related to bundles whose
fibre is an infinite dimensional Clifford algebra.
In this paper we approach the study of these
bundles through
the theory of infinite dimensional principal bundles
whose structure group is the restricted unitary group.

In \cite{CarMur} two of the authors began a related
study: that  of string structures. Our ideas were partly influenced by the
history above. An abstract version of the problem discussed in \cite{CarMur}
is to start with a principal bundle $P$
over a manifold $M$ with structure group $G$.
Let $\hat G$ be a central extension of $G$ by $U(1)$.
Then one can ask
when there exists a principal bundle $\hat P$ with structure group
$\hat G$ such that $\hat P/U(1)=P$ (we call this the extension problem).
Brylinski \cite {Br}
observed that the obstruction to the existence of $\hat P$,
may be identified with a
class in $H^3(M, \Z)$ (Cech cohomology) studied in a different context
by
Dixmier and Douady \cite{Dix}.
Finally in \cite{CM} and \cite{CMM} the connection between
Hamiltonian anomalies, the characteristic classes arising in the
Atiyah-Singer families index theorem and
the Dixmier-Douady class was established.

In this paper we attempt to unify some of
these various manifestations of
the Dixmier-Douady class.
We start by showing (in theorem 4.1) that
if $G$ is simply connected the Dixmier-Douady class
of a principal bundle $P$ is the transgression of the Chern class
of the $U(1)$ principal bundle $\hat G\rightarrow G$.
Next we develop an obstruction theory for the
extension problem showing (Section 5) that it too leads
to the Dixmier-Douady class.

We find that the physical examples discussed above can all be  related to
principal bundles with structure group the restricted unitary group.
  To explain this
 let $H$ be a complex Hilbert
space and $P_+$ an orthogonal projection on $H$
with infinite dimensional kernel and co-kernel.
Denote by $\Ur$ the group of unitary operators $U$ on $H$ such
that $UP_+-P_+U$ is Hilbert-Schmidt and by $\PU$ the projective unitary
group.
The existence of $\Ur$ bundles with non-trivial Dixmier-Douady
class and their relation to the work of Brylinski et al
is covered in Sections 7 and 8.  This is handled by exploiting the
existence of an embedding of the smooth
loop group $L_dG$ of a compact Lie group $G$ into
$\Ur$.
There are  canonical central extensions of both
$L_dG$ (\cite{PreSeg}) and $\Ur$ which are compatible with the embedding
of the former in the latter. Now
Killingback \cite{Kil} argued that the obstruction to extending
a principal $L_dG$ bundle over the space of smooth loops in $M$
  to  a principal bundle having fibre equal to this
extension  transgresses  to  half the Pontrjagin class of $M$.
On the other hand it was shown by
 Brylinski \cite {Br} that the obstruction is the
 Dixmier-Douady class.
Following McLaughlin \cite{Mcl} and \cite{CarMur}
we can prove equality of these
establishing as a corollary the existence of
principal $\Ur$ bundles with non-trivial Dixmier-Douady class.

Our next result concerns the connection between $\Ur$ bundles
and $\PU$ bundles.
 There is a standard inclusion of $\Ur$ into $\PU$ which we
review in Section \ref{res-unitaries}.
Let $P(M,\Ur)$ be a principal $\Ur$--bundle over
$M$. We use the the prefix $\Sigma$ to denote the reduced suspension
of a space and $\Sigma^q$ to denote the suspension isomorphism on cohomology
$$
\Sigma^q: H^q(M,\Z) \cong H^{q+1}(\Sigma M,\Z).
$$
One of our main results (Section \ref{3.18}) is that for $M$ compact,
there is an associated $U(\infty)$-bundle,
 $\Sigma P(\Sigma M,U(\infty ) )$
(an element of $\tilde{K}^1(M)$) over $\Sigma M$ with
$$      \Sigma^3 (D(P)) = c_2(\Sigma P)$$
(the right hand side being the second Chern class of $\Sigma P$).
We deduce from this that
  the structure group of a $\PU$--bundle, $Q$ reduces to
$\Ur$ if and only if there is a $\Ur$ bundle, $P$ whose
Dixmier-Douady
class coincides with that of $Q$.  This happens if and only if
there is a $U(\infty )$--bundle, $\Sigma P(\Sigma M,U(\infty ))$ over
$\Sigma M$ such that
$c_2(\Sigma P)
= \Sigma^3 (D(Q))$.

There are interesting connections between
this paper and a number of other
recent results. For example
another way of viewing the extension of a principal
$G$ bundle to a principal $\hat G$ bundle
is to use the recently introduced notion of a
bundle gerbe \cite{Mur}. In this exposition we
have avoided use of that viewpoint although it
has partly motivated our arguments in section 4 and we discuss
it briefly in Section 12.
The original construction of the Dixmier-Douady class
\cite{Dix} was in connection with bundles of $ C^*$-algebras
with fibre the compact operators
and hence with principal bundles whose fibre is $\PU$.
In the case of principal $\Ur$ bundles
the associated $C^*$-algebra bundles have fibre
the infinite dimensional Clifford algebra. Specifically,
in Section 12 we
associate to any principal $\Ur$ bundle over $M$
a bundle whose fibre is the $C^*$-algebra of the canonical
anticommutation relations (CAR) over $H$ (an algebra isomorphic to
the infinite dimensional Clifford algebra). The vanishing of the
Dixmier-Douady class
allows us to construct an associated Hilbert bundle over $M$ whose
fibre is a representation space for the CAR-algebra (in fact it is
a Fock space) such that the sections of the CAR-bundle over $M$ act
on sections
of the Fock bundle in the obvious way.
Finally, one of the most interesting by-products of our investigation is
the explicit construction in Section 6 of the classifying space of
$\PU$.

\section{Preliminary material on principal $G$-bundles}
We recall some facts about principal $G$ bundles starting with the
definition.  A (topological)
principal $G$ bundle over a topological space $M$ is a triple
$P(M,G)$ where $G$ is a topological group (the structure
group) and $P$ (the total space) and the base $M$ are
topological
spaces with a continuous surjection $\pi \colon
P \to M$.  The group $G$ acts continuously and freely on the right of
$P$
and the orbits of this action are precisely the fibres of the map
$\pi$.  We require that the bundle is {\em locally trivial}
in the sense that
there  is a locally finite cover $\{ U_\alpha \mid \alpha \in A\} $ of
$M$
with the property that if
$P_\a = \pi^{-1}(U_\a)$ then there are homeomorphisms
$P_\a \to U_a \times G$ which send $p$ to
$(\pi(p), s_\a(p))$ and which commute with the action of
$G$ so that $s_\a(pg) = s_\a(p)g$.
Note that the trivial bundle
$M \times G$ is naturally a principal bundle $G$ bundle over $M$ if
we define the obvious right action $(m, h)g = (m, hg)$.
Two principal bundles $P(M, G)$ and $Q(M, G)$ are said to be isomorphic
if there is a homeomorphism $f \colon P \to Q$ commuting
with the $G$ action and the projection map so that the induced
action on $M$ is the identity.
We will be interested in isomorphism classes of principal bundles
which may be classified in two ways that we detail in the next sections.

All that we have said so far
holds also in the category of manifolds and smooth maps
with the corresponding modifications to the
definitions.
In particular many of the principal bundles that we discuss below arise as
the quotient of a Lie group $G$ by a closed subgroup $H$. To show
that $G \to G/H$ is a principal $H$ bundle over $G/H$ one needs to
demonstrate that
this fibration is locally trivial in the topological sense. In all the
cases which arise in this paper both $G$ and $H$ are Banach Lie
groups and the result follows by a theorem of E. Michael (\cite{Mi})
on the existence of local continuous sections for the fibration
$G \to G/H$.

\subsection{Principal bundles and non-abelian cohomology}
Notice that the function $s_\a s_\b^{-1} \colon P_\a \cap P_\b
\to G$ is constant on fibres and hence descends to
define the {\em transition functions} of $P$  with respect to the
 cover by
$$
g_{\alpha\beta }\colon U_\a \cap U_\b \to G.
$$
It is straightforward to check that the  transition functions
$g_{\a\b}$ form a Cech cocycle for the
sheaf $\underline G$ of continuous $G$-valued functions on $M$. It is also
straightforward
to check that if the trivialisations are changed then the cocycle
changes by a coboundary. Hence a principal bundle defines a
class in $H^1(M, {\underline G})$.  Moreover it is possible to
show by the  standard `clutching construction' (see for example \cite{Hus})
that
every cohomology
class arises in this way. We have:

\begin{proposition}
\label{prop1.3} The isomorphism classes of principal $G$ bundles
over $M$ are in bijective correspondence with
the elements of $H^1(M, {\underline G})$.
\end{proposition}

It is important to note that the cohomology space $H^1(M, \underline
G)$
is not a group. It is a pointed set, pointed by the
equivalence class of the identity cocycle which corresponds
under the isomorphism from \ref{prop1.3} to the trivial $G$ bundle.

\subsection{Classifying spaces for principal $G$ bundles}
Another way of describing the isomorphism classes of principal
 bundles is to use classifying spaces. If $f \colon N \to
M$ is a map and $P(M, G)$ is  principal bundle then there is a {\em
 pull-back} bundle $f^*(P)(N, G)$ defined by
 $$
 f^*(P) = \{ (p, n) \colon \pi(p) = f(n)\} \subset P \times N.
 $$
We make $f^{*}(P)$ a topological space or manifold
by its definition as a subspace or submanifold of $P \times N$.
The  action of $G$ is $(p,n)g = (pg, n)$.

 A principal $G$ bundle $EG(BG, G)$ is called a
 {\em classifying space}
for principal $G$ bundles if it has  the property that for any
principal bundle $P(M, G)$ there is a map $f$, unique
up to homotopy, such that $f^*(EG)$ is isomorphic to $P$.
The map $f$ is called a {\em classifying map} for $P$.
A standard fact, see for example, \cite{Hus} is that classifying
spaces exist and are unique up to homotopy equivalence.

It is  sufficient for our purposes to work
in the category of spaces with the homotopy type of a CW--complex,
denoted $CW$ (see, for example, \cite{Spa} pp. 400). Any
map between two CW-complexes whose associated maps on the homotopy
groups are all isomorphisms (a weak homotopy equivalence) is, in
fact, a homotopy equivalence  (\cite{Spa} pp. 405).
For example, differentiable manifolds have the homotopy
type of a CW-complex and $CW$ is closed under the operation of
forming loop spaces.  An extremely useful characterisation of classifying
spaces within the category $CW$
is the fact that a principal $G$-bundle, $P(M,G)$ is a classifying space
if and only if $P$ is weakly contractible (i.e. $\pi_q(P) = 0 $ for all $q$).
Recall Kuiper's Theorem \cite{Kui} which
states that $U(H)$, the full unitary group of a separable Hilbert
space is contractible in the uniform
topology.  This makes $U(H)$ a candidate for the total space of
$CW$--universal bundles.

If $EG(BG, G)$ is a classifying space for $G$-bundles we
may summarise our discussion as:

\begin{proposition}
\label{prop:classifying}
The set of isomorphism classes of principal $G$-bundles
over $M$  is in bijective correspondence with the set of
homotopy classes of maps from $M$ to $BG$.
\end{proposition}

\subsection{Characteristic classes of principal $G$ bundles}
A characteristic class, $c$, for  principal $G$ bundles assigns to
any principal $G$ bundle $P(M, G)$ an element $c(P)$ in
$H^*(M)$, the cohomology of $M$. This assignment is
required to be {\em natural} in the sense that if $f \colon N \to M$
and
$P$ is a $G$ bundle over $M$ then
$$
c(f^*(P)) = f^*(c(P)).
$$
Note that, among other things, this implies that
$c(P)$ depends only on the isomorphism class of $P$.
The results above on classifying spaces give us a complete
characterisation of all characteristic classes. If $c$ is a
characteristic
class we can apply it to $EG$ and obtain an element $\xi = c(EG) \in
H^*(BG)$.
Conversely if $\xi \in H^*(BG)$ then we can define a characteristic
class
by defining $c(P) = f^*(\xi)$ where $f$ is a classifying map for $P$.
So characteristic classes are in bijective correspondence with
the cohomology of $BG$.

\subsection{Associated fibrations}
We shall need to consider other fibrations that arise as {\em
associated fibrations} to a principal bundle. If $P(M, G)$
is a principal bundle and $G$ acts on the left of a  space $X$
then $G$ acts on $P \times X$ by $(p, x)g = (pg, g^{-1}x)$
 and the quotient $(X \times G) / G$
is a fibration over $M$ with fibre isomorphic to $X$.

\section{Changing the structure group}
Let $\phi \colon H \to G$ be a topological group homomorphism.
If $Q(M, H)$ is an $H$ bundle consider the
problem of finding a $G$ bundle $P$ and a
 $\tilde \phi\colon  Q \to P$
such that
\begin{enumerate}
\item $\tilde\phi(Q_{m)}\subset P_{m}$ for all
$m$  in $M$, and
\item $\tilde \phi(qh) = \tilde \phi (q)\phi(h)$
for all $q$ in $Q$ and $h$ in $H$.
\end{enumerate}

This problem can be always solved  in a canonical way.   To define
$P$ we
let  $H$ act on the left of $G$ by $hg = \phi(h) g $
and define $P$ to be the associated fibration to this
action.
The group $H$ acts on $Q \times H$ by $(p, g)g' = (p, gg')$.
The action of $G$ commutes with the action of $H$ and makes
$P$ into a principal $G$ bundle.  We denote it by $\phi_{*}(Q)$.

It is straightforward to show that if we choose local
trivialisations of $Q$ with transition functions $h_{\a\b}$ they
define local trivialisations of $P$ with transition functions
$\phi\circ h_{\a\b}$.  In other words $P$ is
the image of $Q$ under the induced map
$$
\phi \colon H^{1}(M, \underline H) \to H^{1}(M, \underline G).
$$

In terms of classifying spaces we have the following theorem:
\begin{theorem}
\label{th:pushout}
Let  $\phi \colon H \to G$ be a group
homomorphism. Then there is a
map
$$
B\phi \colon BH \to BG
$$
with the property that
if $f \colon M \to BH$ is a classifying map
for an $H$ bundle $Q$ then $B\phi \circ f \colon M \to BG$
is a  classifying map for the  $G$ bundle $\phi_{*}(Q)$.
\end{theorem}
\begin{proof}
This follows from the standard constructions of the
classifying map and the classifying space (see
for example \cite{Hus}).
\end{proof}

More interesting is the `inverse' problem to this. If $P(M, G)$
is a principal bundle can we find a principal $H$ bundle
$Q$ such that $\phi_{*}(Q)$ is isomorphic to
$P$?  A number of ways of deciding when this is possible
are known.

First, in terms of  C\'ech cohomology:
a bundle $Q$ exists if the bundle $P(M, G)$ lies in the image
of
$$
\phi\colon H^{1}(M, \underline H ) \to H^{1}(M, \underline G).
$$
Second, in terms of classifying spaces we have
\begin{theorem}
\label{th:extending}
Let $\phi \colon H \to G$ be a group
homomorphism. Then if  $f \colon M\to BG $ is a
classifying map for $P$ then a $Q$ bundle $H$ exists  with
$\phi^{*}(Q) \simeq P$ if and only
if $f$ lifts to a map $\hat f \colon M \to BH$ such
that $B\phi \circ \hat f = f$.
\end{theorem}
\begin{proof}
This follows from Theorem \ref{th:pushout}.
\end{proof}

The third method, which will be explained in the
examples below,  is to formulate the problem
as that of finding a section of a fibration and
to  employ obstruction theory.

We are interested in  two particular cases of this general problem:
\begin{enumerate}
\item $H$ is a closed Lie subgroup of $G$
\item $\hat G \to G$ is a central extension with kernel $U(1)$.
\end{enumerate}
In the first of these cases we say that the structure group $G$
{\em reduces} to $H$ and in the second that it {\em lifts} to $\hat
G$.

\subsection{Reducing the structure group}
Let $H$ be  a closed Banach Lie subgroup of
a Banach Lie group $G$. If $Q(M, H)$ is
a principal bundle with a bundle map from $Q(M, H)$ to $P(M, G)$
then it identifies $H$ with its image inside $P$.
This image is a {\em reduction} of $P$ to $H$. That is, it is  a
submanifold of $P$ which is stable under $H$ and forms,
with this $H$ action, a principal $H$ bundle over $M$. It is clear
that the problem
of reducing $P$ to $H$ is equivalent to the problem of finding a reduction
 to $H$.
 Given a bundle $P(M, G)$, consider a fibre $P_{m}$.  A reduction
of $P$ involves selecting an $H$ orbit in $P_{m}$ for
 each $m$.
The set of all $H$ orbits in $P_{m}$ is $P_{m}/H$ and a reduction
of $P$ therefore corresponds to a section of the fibering  $P/H \to M$
whose fibre at $m$ is $P_{m}/H$.

Applying this to the classifying space of $G$  we see that
$EG \to EG/H$ is a principal $H$ bundle with contractible total space
and hence a classifying space for $H$.
The map $H \subset G$ induces a map $BH \to BG$ which
under these identifications is the map $EG/H \to BG$.
It is now straightforward to show that the following
theorem holds.

\begin{theorem}
\label{2.3thm}
Let $P(M,G)$ be a principal $G$--bundle with classifying map
$f \colon M \to BG$ then the following  conditions are equivalent
to the structure group of $P$ reducing to $H$:

\begin{enumerate}

\item  The fibration $P/H \to M $ has a global section.
\item The classifying map, $f$, has a lift, $\hat{f}$, to $BH = EG/H$.

\end{enumerate}

If, in addition, $H$ is normal in $G$, then a final equivalent
condition is
$\rho [P] = 0$ where $\rho $ is the map in first cohomology induced by
the canonical projection $G\longto  G/H$
$$               \rho : H^1(M,\underline{G} ) \to
H^1(M,\underline{G/H} ).$$
\end{theorem}
\begin{proof}
(1) Defining a reduction of $P$ to $H$ means   picking out,
for each $m$ in $M$  an orbit
of $H$ inside  $P_{m}$ or equivalently an element of $P_{m}/H$.
But the latter defines a section of $P/H$.

(2) Theorem \ref{th:extending}.

(3) If $P$ has a reduction to $H$ then we can always
choose our local trivialisations so that the transition
functions take values in $H$. Hence $\rho(P) = 0$.
Conversely if the transition functions are $g_{\a\b}$
and $\rho(P) =  0$ then we must have
$$
g_{\a\b} = g_{\b}h_{\a\b}g_{\a}^{-1}
$$
where $g_{\a} \colon U_{\a} \to G$ and
$h_{\a\b} \colon U_{\a}\cap U_{\b} \to H$. Let the  transition
functions be defined by local trivialisations
$p \mapsto (\pi(p),  s_{\a}(p))$  so that
$g_{\a\b}\circ \pi  = s_{\b}s_{\a}^{-1} $. If we
modify these by letting $s'_\a = s_\a g_\a$ and
$s'_\a = s_\b g_\b$ then we find that  the
new transition functions are $h_{\a\b}$ as required.
\end{proof}

Before  we can apply Theorem \ref{2.3thm} usefully we need
the obstructions to lifting
maps from the base space of a fibre bundle to the total space
({\it {loc cit}} Steenrod
pp. 177 --- 181).  Briefly,  assume
that $M$ is a $CW$ complex and that we are trying to lift a map
$f: M \to B$ to the total space of the fibre bundle $\pi_{E} \colon E
\to B$ with
fibre $F$ such that
the lift, $\Hat{f}$ satisfies $f =\pi_E\composition \Hat{f}$.  We
define
$\hat{f}$ over the zero
skeleton of $M$ by lifting $f$ arbitrarily.  Extending over the
1--skeleton of $M$ is only a problem if the fibre, $F$, is not
connected.
In general, there is no difficulty in extending a map from the
$n$-skeleton to the $(n+1)$-skeleton of $M$ if $\pi_n(F)$ is zero.
We will be interested in the case that $F$ has non-vanishing
homotopy only in one dimension, that is, it is an Eilenberg-Maclane space.
Recall that if $A$ is a group and $n>0$ then we denote by
$K(A,n)$ the Eilenberg-Maclane space whose only non-vanishing homotopy
in a dimension greater than zero occurs in dimension $n$
where $\pi_n(K(A,n))=A$. In this case the general theorem from \cite{Whi}
 page 302 becomes:

\begin{theorem}
\label{th:obstruction}
Let $f: M \to B$ be a continuous map where $M$ is a $CW$
complex and let $\pi_{E} \colon  E \to B $ be a fibration over $M$ with
fibre $F = K(A,n)$ (i.e. an Eilenberg MacLane space)
with $n >0$ and $A$ abelian.  Then
there exists a cohomology class, ${o(}f,E) \in
H^{n+1}(M, \cal{A})$,
which depends only on the homotopy class of $f$ and which has the property
that $f$  has a lift, $\Hat{f}\colon  M \to E$ if and only if
$ o(f,E) = 0$.

Moreover if  $g: M' \to M$ is continuous, then
$$  o(f\circ g,E)
= g^*(o(f,E)) \in H^{n+1}(M', \cal{A})). $$
\end{theorem}

\begin{note}
Notice that it suffices to define $o(\id, E)$ where
$\id \colon B \to B$ is the identity map. Then
$o(f, E) = f^*(o(\id, E))$.
\end{note}

\begin{note}
\label{note(c)}
We use the notation $H^{n+1}(M, \cal{A})$
to denote the fact that the cohomology may takes values,
not  simply in $\pi_n(F)=A$ but in a possibly twisted $A$
bundle over $B$.  However, when this bundle is trivial we recover
standard cohomology and this is the case precisely when the action of
$\pi_1(B)$ on the fibre is trivial.  Fibrations of this sort may be called
 principal
$K(A,n)$-fibrations and it is easy to check that the pull-back of a principal
$K(A,n)$-fibration is itself a principal $K(A,n)$-fibration.
It follows that when
$K(A,n)$ is realised as a topological group, principal $K(A,n)$-bundles are
 principal
$K(A,n)$-fibrations (since $\pi_1(BK(A,n)) = 0$).
\end{note}

The following lemma allows one to compute the homotopy groups of the fibre of
the map $B\phi: BH \to BG$ in the case that $\phi$ is an inclusion.

\begin{lemma}
\label{lma:HintoG}
Let $i: H \inclusion G$ be an inclusion of topological groups. Then there is a
commutative diagram of homotopy groups for all $q  \geq 0$.

$$
\begin{CD}
1 @>>>  {\pi_q (BH)}   @>{\delta}>>   {\pi_{q-1} (H)} @>>> 1\\
@VVV    @V{Bi_{*,q}}VV                @V{i_{*,q}}VV    @VVV \\
1 @>>>  {\pi_q (BG)}   @>{\delta}>>   {\pi_{q-1} (G)} @>>> 1
\end{CD}
$$
\end{lemma}

\begin{proof}Setting $B:=(EH \times G)/H = B(BH,G,Bi)$ let $Bi'$ be the bundle
morphism
$Bi':B \to EG$ covering $Bi$ and let $I$ be the obvious bundle morphism
$I:EH \to B$ covering $id_H$.  Then $Bi' \composition I:EH \to EG$ is a
bundle morphism covering $Bi$.  The commutative diagram above is just the
commutative diagram of the long exact sequences of the fibrations $EG(BG,G)$
and $EH(BH,H)$ with the map of fibre bundles $Bi' \composition I$
(including the weak contractibility of $EG$ and $EH$).
\end{proof}

\subsection{Obstruction and transgression}
\label{obs and trans}
Recall the spectral sequence of a fibration \cite{MosTan}. If
$E \stackrel{\pi}{\to} B$ is a
fibration with fibre $F$ there is a spectral sequence with
$$
E_2^{p, q} = H^p(B, H^q(F, \Z))
$$
converging to a grading of  the total cohomology of $E$.
If $H^1(F, \Z) = 0$ then the differential $d_3$ of this spectral sequence
defines
a map
$$
\tau = d_3:  H^2(F,\Z) \to H^{3}(B, \Z)
$$
called the {\em transgression} \cite{MosTan}.
Note that this is a different
transgression from that mentioned in the introduction.

A useful fact we will use later is
\begin{proposition}
\label{pr:trans}
If $[x] \in H^{3}(B, \Z)$ and
$\pi^*(x) =  d y$ for a two  class $y$ on $E$ then
 $$\tau([y_{|F}]) = [x].$$
\end{proposition}
\begin{proof}
(\cite{MosTan} page 81.)
\end{proof}

Then we have
\begin{theorem}
\label{th:obs_trans}
Let $f: M \to B$ be a continuous map where $M$ is a $CW$
complex and let $\pi_{E} \colon  E \to B $ a fibre bundle over $M$ with
fibre $F$.
Suppose further that $\pi_2(F) = \Z $ is the only non-vanishing
homotopy group.  Let $\mu$ generate $H^2(F, \Z) = \Z$.
Then $o(\id, E)$ is the transgression of $\pm \mu$.
\end{theorem}
\begin{proof}
See \cite{MosTan} pages 103 and 109.
\end{proof}

\section{Extending the structure group}
Let
\begin{equation}
\label{3A}
1 \to U(1) \to \hat{G} \stackrel{\rho}{\to} G \to 1
\end{equation}
be a short exact sequence of Lie groups with $U(1)$ central.
If $P(M, G)$ is a principal bundle we are interested in the
problem of finding a {\em lift} of $P$ to a $\hat G$ bundle
$\hat P$ over $M$.  We shall present two methods of
defining a characteristic class: the Dixmier-Douady class
and the obstruction class, both of which are
obstructions to finding such a lift. We then
show that they are, in fact, equal.

\subsection{The obstruction class}

\begin{proposition}[\cite{CoqPil}]
We can realise $B\Hat{G}$ as a principal $BU(1)$-bundle over $BG$.
\end{proposition}
\begin{proof}
Steenrod \cite{Ste1} showed that
Milgram's realisation of the classifying space makes $E$ a functor >from the
category of topological spaces and continuous homomorphisms to itself.
In fact we have the following commutative diagram where the vertical arrows
are the inclusion of a fibre.
$$
\begin{CD}
1 @>>> U(1) @>{i}>> \hat{G} @>{\rho}>> G @>>> 1 \\
@VVV @VVV @VVV @VVV @VVV \\
1 @>>> EU(1) @>{Bi}>> E\hat{G} @>{B\rho}>> EG @>>> 1 \\
\end{CD}
$$
Functorality allows us to move from $U(1)$ central in $\hat{G}$ to
$EU(1)$ central in $E\hat{G}$ and thus $U(1)$ is normal in $E\hat{G}$.
Since we have a closed inclusion,
$U(1) \inclusion \Hat{G}$, $E\hat{G} /U(1)$ is a realisation
of $BU(1)$ as a topological group.  Moreover,  $E\hat{G}/U(1)$ is a principal
$G$-bundle over $B\hat{G}$ with $G$ canonically identified as a subgroup.
We may form the associated bundle
$ B := (  EG \times E\hat G/N)/G \to BG$ which is a principal
$E\hat{G}/U(1)$-bundle over $BG$.  However,
we may also project $B$ onto $B\hat G$ and the fibre is
$EG$ which is contractible. So $B$ has the homotopy type
of $B\hat G$ and hence is another realisation of $B\hat G$
proving the result.
\end{proof}

{}From Theorem \ref{th:extending}
we see that if $P(M,G)$ is a principal $G$--bundle with classifying
map $f \colon M \to BG$ then
 $P$ lifts to  $\hat{G}$
if and  only if  there is a lift of $f$ to $B\hat{G}$.
To find when such a lift occurs we can use Theorem
\ref{th:obstruction} from
obstruction theory. The classifying space $BU(1)$ is an
Eilenberg-Maclane space whose only non-vanishing homotopy
is $\pi_{2}(BU(1)) = \Z$.  Since we have realised $B\hat{G}$ as a
principal $BU(1)$-bundle it follows from \ref{note(c)} it  follows that
there is no twisting in the co-efficient group and that the obstruction to
lifting $f$ is a class $O(f) \in H^{3}(M, \Z)$.  The
results of \ref{th:obstruction} imply that
this defines  a characteristic class
in $ H^{3}(M, \Z)$. To get an exact normalisation
for this class we choose the generator $\mu \in  H^2(BU(1),\Z)$
to be the Chern class and then define $O(P) = f^*(\tau(\mu))$.

\subsection{The Dixmier-Douady class}
Because  $U(1)$ is central in $\hat G$ it is possible to
show that there is a short exact sequence of {\em pointed sets}
\begin{equation}
\label{eq:cent-seq}
H^{1}(M, \underline U(1)) \to H^{1}(M,\underline{ \hat G})
 \to H^{1}(M, \underline G)  \stackrel{\delta}{\to} H^{1}(M,
 \underline U(1)).
\end{equation}
The definition of an exact sequence of pointed sets is that
if $X$, $Y$ and $Z$ are sets with points $x$, $y$ and $z$ and
\begin{equation}
X \stackrel{f}{\to} Y \stackrel{g}{\to} Z
\end{equation}
is a sequence of pointed maps (that is $f(x) = y$ and $g(y) = z$)
then this sequence is exact at $Y$ if $f(X) = g^{-1}(z)$. This
clearly agrees with the  definition for groups if the point
of a group is the identity.

The map $\delta$ is defined as follows. Choose
a Leray cover $\{U_\a\}$ and local sections $s_\a \colon U_\a \to P$.
Then the transition functions of the bundle are defined by $s_\a =
s_\b g_{\a\b}$.
We can lift these to maps
$$
\hat g_{\a\b}\colon U_\a \cap U_\b \to \hat G.
$$
Of course these may not be transition functions for a $\hat G$ bundle.
Their failure  to be so is measured by the cocycle
$$
e_{\a\b\c} = \hat g_{\b\c}\hat g^{-1}_{\a\c}\hat g_{\a\b}
$$
which takes values in $U(1)$. Because $U(1)$ is central
it can be shown  that $e_{\a\b\c}$ defines a class
in $H^2(M, \underline U(1))$ which vanishes precisely
when we can lift the bundle $P$ to $\hat G$.

We can use  the short exact sequence of groups
$$
0 \to \Z \to \underline \R \to \underline U(1) \to 0
$$
to define an isomorphism
$$
H^{2}(M, \underline U(1)) \simeq H^{3}(M, \Z).
$$
The result of applying this
isomorphism to $e_{\a\b\c}$ defines a characteristic class
$D(P) \in H^{3}(M, \Z)$ called the Dixmier-Douady class. Explicitly
if we choose $w_{\a\b\c}$ so that $e_{\a\b\c} = \exp(2\pi i
w_{\a\b\c})$
then  the Dixmier-Douady class has a representative
\begin{equation}
	d_{\a\b\c\d} = w_{\b\c\d} - w_{\a\c\d} + w_{\a\b\d} + w_{\a\b\c}.
	\label{eq:dw}
\end{equation}

Note that if $p$ is a point in the fibre $P_m$
above $m$ then there is a homeomorphism $G \to P_m$
defined by $g \mapsto pg$. If $G$ is connected then changing
$p$ gives a homotopic homeomorphism and hence there is
a unique identification of the cohomology of $G$ with
the cohomology of $P$.
We want to prove
\begin{theorem}
\label{th:G_trans}
Let $P \to M$
be a principal $G$ bundle with $G$ one-connected.
Let $\hat G \to G$ be a central extension of $G$ by $U(1)$.
Let $[\mu]$ be the cohomology class (the Chern class of $\hat G \to
G$) that the central extension defines on $G$
and hence also on any fibre of $P \to M$.
Then the transgression of $[\mu]$ is the Dixmier-Douady class
of the bundle $P \to M$.
\end{theorem}
\begin{proof}
To do this we need an alternative definition of
transgression from \cite{Spa}. We start with  $\pi \colon P \to M$
as above and $P_{m_0} = \pi^{-1}(m_0)$ the fibre above some fixed $m_0$.
Then there are homomorphisms:
$$
H^q(P_{m_o}, \Z) \stackrel{\delta}{\rightarrow} H^{q+1}(P, P_{m_0}, Z)
\stackrel{\pi^*}{\leftarrow}  H^{q+1}(M, \{{m_0}\}, \Z)
\stackrel{j^*}{\rightarrow} H^{q+1}(M, \Z).
$$
The transgression is then the map
$$
\tau\colon \delta^{-1}(\text{Im}(\pi^*) \rightarrow H^{q+1} (M , \Z)
/j^*(\text{Ker}(\pi^*).
$$
defined by $\tau(u) = j^*(\pi^*)^{-1} \delta(u) $ where
$u \in H^q(F,  \Z)$ is such that $\delta(u) \in p^*(H^{q+1}(B, b_0,
\Z))$.

For our purposes it is most useful to realise the cohomology
here using C\'ech cocycles. Recall that if $X$ is a topological
space and $A$ is a subspace we can define the
relative integral cohomology $H(X, A, \Z)$ as follows.
We take a cover $\U$ of $X$ and a subcover $\U' \subset \U$
of $A$ and consider the induced map of complexes of  C\'ech cocycles
for the group $\Z$:
$$
C^p(X, \U) \to C^p(A, \U').
$$
We define $C^p((X, A), (\U, \U'))$ to be the
kernel of this map and define
$$
H^p((X, A), (\U, \U'), \Z)
$$
to be the cohomology of the complex
$C^p((X, A), (\U, \U'))$. The cohomology group $H^p(X, A, \Z)$ is
now defined in the usual way by taking the direct limit as the
covers are refined.

In the particular case of interest choose a cover $\U$ of
$M$ with respect to which the Dixmier-Douady class can be represented
by a cocycle  $d_{\a\b\c\d} \in C^3(M, \U)$ as in \eqref{eq:dw}.
Choose $\a_0$ so that
 $m_0 \in U_{\alpha_0}$ and let $\U' = \{ U_{ \a_0} \}$.
Then the restriction of any cocycle in $C^3(M, \U)$
to $C^3(m_0, \U') $ is automatically zero.

Consider the transition functions
$g_{\a\b}$. The pullback of $g_{\a\b}$ to $P$ is trivial
because it satisfies
 $$
\pi^*g_{\a\b} = \sigma_\b\sigma_\a^{-1}
$$
where $\sigma_\a (p ) $ is defined by $\sigma_\a(s_\a(x)g) = g$.

Restricted to any fibre the maps $\sigma_\a \colon P_m \to
G$ are homeomorphisms. Cover $G$ by open sets $V_a$
over which $\hat G \to G$ has transition functions $h_{ab}$
relative to local sections $r_a \colon V_\a \to \hat G$.
Then we can use the maps
$$
\begin{array}{ccc}
\pi^{-1}(U_\a) & \to & U_\a \times G \\
p  &  \mapsto &(\pi(p), \sigma(p)
\end{array}
$$
to pull the  $V_a$ back to $P$ to define open sets $W_{(\a, a)}
\subset \pi^{-1}(U_\a)$. The cover
 $\W = \{ W_{(\a, a)}\}$ is a refinement of the cover $\{\pi^{-1}(U_\a)\}$.
If $\rho_{\a_1, \a_2, \dots, \a_d}$ is a cocycle for $\{\pi^{-1}(U_\a)\}$
we denote by $\rho_{(\a_1, a_1), (\a_2, a_2), \dots, (\a_d, a_d)}$ its
restriction to
 $\{ W_{(\a, a)}\}$.

  In particular consider the $G$ valued cocycle
 $\sigma_{(\a, a)}$. This can be lifted to $\hat G$ by defining
 $\hat \sigma_{(\a, a)} = r_a \circ \sigma_{(\a, a)}$. Then
$\pi^*\hat g_{(\a, a)(\b, b)}$ and $\hat \sigma_{(\b, b)}\hat \sigma_{(\b,
b)}^{-1}$ are both lifts of $\pi^*g_{(\a, a)(\b, b)}$ so that
we must have
\begin{equation}
\hat \sigma_{(\beta, b)}\hat \sigma_{(\a, a)}^{-1} h_{(\a, a)(\b, b)}
= \hat \pi^*\hat g_{(\a, a)(\b, b)}
\label{eq:v}
\end{equation}
for a cocycle
$$
h_{(\a, a)(\b, b)} \colon U_{(\a, a)}\cap U_{(\b, b)} \to U(1).
$$
Hence we have
$$
\pi^*e_{(\a, a)(\b, b)(\c, c)} =
h_{(\b, b)(\c, c)}h^{-1}_{(\a, a)(\c, c)}h_{(\a, a)(\b, b)}.
$$
Letting $h_{(\a, a)(\b, b)} = \exp(2\pi i v_{(\a, a)(\b, b)})$ gives
\begin{equation}
\pi^*w_{(\a, a)(\b, b)(\c, c)} =
 v_{(\b, b)(\c, c)} -  v_{(\a, a)(\c, c)} +
  v_{(\a, a)(\b, b)} + n_{(\a, a)(\b, b)(\c, c)}
\label{eq:n}
\end{equation}
for $n_{(\a, a)(\b, b)(\c, c)} $ some integer valued co-cycle.
Finally we deduce that
\begin{equation}
\pi^*d_{(\a, a)(\b, b)(\c, c)(\d, d)} =
 n_{(\b, b)(\c, c)(\d, d)} -  n_{(\a, a)(\c, c)
 (\d, d)} + n_{(\a, a)(\b, b)(\d, d)}
- n_{(\a, a)(\b, b)(\c, c)}.
\label{eq:dn}
\end{equation}

Consider now the cohomology on the fibre $P_{m_0}$.
We define a cover $\W'$ which covers $P_{m_0}$ by
$$
\W' =  \{ W_a = W_{(\a_0, a)} \}.
$$
 We make corresponding
notational changes to indicate restriction of cocycles from $\W$ to $\W'$.
For example
the restriction of $n_{(\a, a)(\b, b)(\c, c)}$ is
 $n_{abc} = n_{(\a_0, a)(\a_0, b)(\a_0, c)}$. We
then have from equation \eqref{eq:n}
$$
0 = \pi^*w_{(\a_0, a)(\a_0, b)(\a_0, c)} =
 v_{bc} - v_{ac} + v_{ab} + n_{abc}
$$
so that
\begin{equation}
n_{abc} = -v_{bc} + v_{ac} - v_{ab}
\label{eq:nv}
\end{equation}
Using equation \eqref{eq:v} we see that
$$
\hat\sigma_b \hat \sigma_a^{-1} =
\hat \sigma_{(\a_0, a)}\hat \sigma_{(\a_0, b)}^{-1}
=  \exp( - 2\pi i v_{ab}).
$$
Finally note that $\hat\sigma$ is defined
by $\sigma_{(\a, a)} = r_a \circ
\sigma_\a{|U_{(\a, a)}}$ so that
$$
 \exp( - 2\pi i v_{ab}) = (r_b r_a^{-1}) \circ \sigma_{\a_0}
 $$
where  $r_a r_b^{-1}$ are the transition functions
of $\hat G \to G$.

Finally we can
calculate the transgression of the Chern class.
It follows >from \eqref{eq:nv} that the Chern class in $H^2(P_{m_0}, \Z)$
is represented by the cocycle $n_{abc}$. We want to
apply the coboundary map in relative cohomology to this
to obtain a class in $H^3(P, P_{m_0}, \Z)$.
 We do this by first extending
$n_{abc}$ to a class on all of $P$ and then applying the
C\'ech coboundary to it. But we obtained $n_{abc}$ by restricting
$n_{(\a, a)(\b, b)(\c, c)}$ so this is an obvious extension
and then \eqref{eq:dn} shows that if we apply the C\'ech
coboundary to $n_{(\a, a)(\b, b)(\c, c)}$ we obtain
the class $\pi^*d_{(\a, a)(\b, b)(\c, c)(\d, d)}$
which is the pullback of the Dixmier-Douady class as required.
  \end{proof}

\section{Equality of the two classes}
This Section is devoted to the proof of the
following fact.
\begin{theorem} The obstruction  and the Dixmier-Douady
classes are equal.
\end{theorem}
\begin{proof}
Notice first that the universal
bundle for $U(1)$,  $EU(1) \to BU(1)$, can be
realised as $E\hat G \to E\hat G / U(1)$. Also
we have that $G$ acts on $E\hat G / U(1)$ and
hence we can form the associated fibration
$$
(EG \times E\hat G/S^1) /G \to BG.
$$
The fibres of this are therefore $BU(1)$.
Notice also that if we project onto $B\hat G$ that
fibering has contractible fibres and $(EG \times E\hat G/S^1) /G$
is homotopy equivalent to $B\hat G$.

Consider the diagram
$$
\begin{array}{ccc}
\hat G & \rightarrow & E\hat G \\
\downarrow  &  & \downarrow \\
 G & \rightarrow & E\hat G / U(1).
\end{array}
$$
It follows that the bottom arrow must be the classifying map.
Let $\mu$ be the generator of $H^2(BU(1), \Z)$. Let
$f $ be the classifying map. Then $f^*(\mu)$ is the class
of the bundle $\hat G \to G$.

We now have a  commuting diagram of fibrations:
\begin{equation}
\begin{array}{ccccc}
 EG && \stackrel{\tilde f} {\rightarrow} && B\hat G \\
&\searrow  &  & \swarrow &\\
& & BG & &
\end{array}
\label{eq:fibrations}
\end{equation}
where the map $\tilde f$ restricted to fibres
is the classifying map $f$.

Let us denote by $[\mu]$ the class
on a fibre of $B\hat G \to BG$ which is the
fundamental class in $H^2(BU(1), \Z) = \Z$.
Then by Theorem \ref{th:obs_trans} we have that
this transgresses to the obstruction class.
Also  by Theorem \ref{th:G_trans}
the class $\tilde f^* ([\mu])$ restricted to a fibre
transgresses to the Dixmier-Douady class.
But for the  commuting diagram \eqref{eq:fibrations} of fibrations
the  transgression maps  will commute with
$\tilde f^*$ and hence the
obstruction and Dixmier-Douady class coincide.
\end{proof}

\section{The classifying space of the projective unitaries}
\label{3.9section}
Given the importance of $\PU$ and principal $\PU$--bundles in the
following theory
we remark that there is a simple construction of a  $B\PU$ which is
a homogenous, infinite dimensional smooth manifold and will allow
us to obtain a $BG$ when $G \inclusion \PU$ is a closed embedding
of Banach Lie groups.
Throughout this section all groups are equipped with their
natural Banach Lie group topologies (in the case of $\PU$ this arises from
 the norm topology on the unitary group).

\begin{proposition}
\label{3.10prop}
There exists a closed inclusion of $QU(H) = U(H)/U(1)$ in
 $\U$, the
unitary group of the Hilbert space of Hilbert--Schmidt operators
$\mathcal T$ on
$H$ ($\U$ is equipped with the norm topology).
\end{proposition}
\begin{proof}
 Given $[a] \in \PU$, choose a representative $a \in U$.
Then define
$$               i: \PU \to \U$$
$$                [a] \mapsto  Ad(a) $$
where
$$               Ad(a): {\mathcal T} \to  {\mathcal T} $$
$$                   t \mapsto  a.t.a^*$$
Clearly $i$ is well defined and is
injective. To prove continuity of $i$
considering any convergent sequence $([a_n])_{n=1}^\infty \to [1]$ in
$\PU$.  By taking
$n$ large enough we may assume that the $[a_n]$ lie in a neighbourhood
 over which $U(\PU,S^1)$ is locally trivial.  Hence we may
assume that there is a sequence $(a_n)_{n=N}^\infty \to 1$ in $U$.
Then it is straightforward to see that
$\| Ad(a_n) - Ad(1) \|_{B(\cal{T})} \to 0$ as $n \to \infty$.
To see that the image of $i$ is closed consider a sequence
$i([a_n]) \to b$ where
$b \in \U_u$.  Define a $*$--automorphism of $\mathcal {T}$ by
$$		b'(t) = \lim_{n \to \infty } Ad(a_n)t.$$
One can verify that $b'$ is a $*$--automorphism of ${\mathcal T}$.
  Since ${\mathcal T}$ is uniformly
 dense in ${\mathcal{K}}(H)$, the compact operators on $H$,
$b'$ defines a $*$--automorphism of ${\mathcal {K}}(H)$ and is thus of the
form $Ad(a)$ for some $a \in U(H)_u$.  Hence $b = Ad(a) =
i([a])$ and the image of $i$ is closed.

Finally, to see that $i$ defines a homeomorphism we begin with the metric,
$\rho$,
which defines the topology on $\PU(H)$.
$$\rho([a],[b]) = Inf(\lambda \in S^1) \| a - \lambda .b \|
_{B(H)}$$
Now let $\Theta_{u,v}$ be the rank one operator
$\Theta_{u,v}: H \to H$ given by
$ w \mapsto (v,w)u$
Then the  map
$ u\otimes v \mapsto\Theta_{u,v}$
extends to an isomorphism of $\overline{H}\otimes H$
 with $\mathcal{T}$. Here the bar denotes the
complex conjugate Hilbert space. The operator $Ad(a)$
becomes $\bar{a}\otimes{a}$ where $\bar a$ denotes
the action of $a$ on the conjugate space.
To prove our result it suffices to work in a neighbourhood
of the identity in $\U$. Now for $\bar{a}\otimes{a}$ to be close
 to the identity operator the spectrum of $a$ must contain a gap (for if
the spectrum is the whole circle then it is not possible for
 $\bar{a}\otimes{a}$ to be close to the identity).
That being the case we can assume $-1$ is not in the spectrum
of $a$ by multiplying by a phase if necessary.
Assume we have a sequence
 $a_n \in U(H)$ with
$$
\| Ad(1)-Ad(a_{n}) \|_{B({\mathcal T})} \to 0.
$$
Then there is a sequence of self adjoint operators $K_n$ on
 ${\mathcal T}$ with $a_n=\exp(iK_n)$ and the spectrum
of $K_n$ is $[\gamma_n, \delta_n]\subset [-\pi,\pi]$.
In fact we may assume
 $$
 \gamma_n=\inf_{(\| u \|_H = 1)}\{(u,K_n u)\},$$
  $$
  \delta_n=\sup_{(\| u \|_H= 1)}\{(u,K_n u)\}.
   $$
Then
\begin{align*}
||Ad(a_n)-Ad(1)||&= \sup\{|\exp i(\lambda-\mu)-1|\
 |\ \lambda,\mu\in [\gamma_n, \delta_n]\} \\
&=  \exp i(\delta_n-\gamma_n) - 1
\end{align*}
On the other hand
\begin{align*}
\inf_\lambda ||a_n-\lambda 1|| &= |\exp[i(\delta_n-\gamma_n)/2]-1|\\
&=||Ad(a_n)-Ad(1)||.
\end{align*}

Thus, if $\| Ad(1) -Ad(a_n) \| _{B(H)} \to 0$ as $n \to \infty$
then $\rho([1],[a_n]) \to 0$.
Hence $i^{-1}:i(\PU(H) \to \PU(H)$ is continuous and
thus $i$ is a homeomorphism.

\end{proof}

This result shows that $\PU$ is a Banach Lie subgroup of $\U$.
The contractibility of $\U$ (Kuiper's theorem)
means that (after identifying $i(\PU)$
and $\PU$), we have that
$$
\U(\U/\PU,\PU)
$$
is a
locally trivial (by \cite{Mi}) universal
$\PU$--bundle and that $\U/\PU$
is a $B\PU$.  More generally, if $G$ is a closed sub-Banach Lie Group of
$\PU$, then $\U(\U/G,G)$ is a universal $G$-bundle.

\section{String structures}
\label{3.13section}
We start
with a  principal $G$--bundle, $P(M, G)$  where $G$ is a compact
Lie group and form the bundle
$L_dP(L_dM,L_dG,Lf)$ where, in general, $L_dM$ denotes the space of
differentiable loops into a finite dimensional manifold $M$.
It is well known (\cite{PreSeg} Ch 6)
that $L_dG$ has a canonical central extension by $S^1$,
$\widehat{L_dG}$, induced from an embedding of $L_dG$ in the restricted
unitary
group  which in
turn embeds in the projective unitaries of a second Hilbert space
$H_\pi$.   Henceforth $U$ and $\PU$ will refer
respectively to
the unitaries and projective unitaries over $H_\pi $.
$$               L_dG \inclusion  \Ur  \inclusion \PU(H_\pi ),$$
$$      \widehat{L_dG}(L_dG,S^1) = i^*U({H}_\pi )(\PU({\mathcal
H}_\pi ),S^1),$$
$$        [\widehat{L_dG}]\text{ generates }H^2(L_dG,\integers).$$
The idea of a string structure
arises as follows. Starting with a principal
$SO(n)$--bundle,
$$P(M,SO(n),f)$$
 ($n>2$), which is usually the frame bundle
of a tangent bundle, $TM$,
and which has a $Spin(n)$ structure $Q(M,Spin(n),\hat{f})$ with classifying
map $\Hat{f})$ one
 forms the loop bundle
$$L_dQ(L_dM,L_dSpin(n),L_d\Hat{f}).
$$
The bundle  $P$ is said to
have a {\em string structure} if and only if  the structure group of $L_dQ$
extends to
$\widehat{L_dS}pin(n) $.
  Of course, the Dixmier-Douady class of
$L_dQ$, $D[L_dQ]$, is the obstruction to the existence of a string
structure.
Killingback proposed that twice $D[L_dQ]$ was in fact the transgression
of the Pontryagin class of $P$.  Since then MacLauglin \cite{Mcl} and
Carey and Murray \cite{CarMur} have produced rigorous proofs of
Killingback's result.

\subsection {Loop spaces, groups and bundles}
Henceforth, let $X$ be a topological space, $H$ a topological
group, $M$ a finite dimensional manifold and $G$ a compact
Lie group.  By $(\Omega_d M,m_0)$ we denote the based, differentiable loops
into $M$.
$$     \Omega_d(M,m_0) := \{\gamma \in L_d(M): \gamma(0)=\gamma(1)=m_0\}$$
When the base point is unimportant we shall suppress it.
$L_cX$ and $\Omega_cX$
shall denote the spaces of continuous loops and continuous
based loops respectively, both with the compact open topology.
Whereas, $L_sM$ and $\Omega_sM$
shall denote the loop spaces used by Carey and Murray \cite{CarMur}
consisting of free or based continuous loops differentiable except perhaps
at $m_0$.   $L_dX$ and $L_sX$ have the structure of differentiable
Frechet manifolds when given the Frechet topology (see \cite{CarMur}).
Moreover, in the case where the spaces are groups,
$L_c H$ and $L_{s,d} G$ have, respectively, the structure of a
topological group or a Lie group under pointwise multiplication of loops.
(When considering based loops into a group, the base point is taken to
be the identity of the group.)  We shall next show that all three loop
spaces are homotopic and hence
they share many properties.  When dealing with facts and properties
equally applicable to either the differentiable, piecewise
differentiable or continuous loops
we shall drop the subscripts and use $LX$ and $\Omega X$ where it is
understood that X is a manifold if the loop functor in question is any
of $L_d$, $L_s$, $\Omega_d$ or $\Omega_s$.

\begin{proposition}
\label{3.14prop}
Let $M$ be a differentiable manifold of finite or infinite dimension, then
$\Omega _cM$, $\Omega _sM$ and $\Omega _d M$ have the same homotopy type.
\end{proposition}
\begin{proof}
We shall show that the obvious inclusions $i: \Omega _dM \inclusion \Omega
_sM$,
$j: \Omega _s M \inclusion \Omega _c M$ and $j\composition i$ are
 weak homotopy equivalences.  Then, since  $\Omega _c M$, $\Omega _s M$ and
$\Omega _d M \in CW$, it will follow that they are of the same homotopy
type.  Firstly, we start with some standard notation and the case of
 $j\composition i$:
$$
    I^n  := \{ (y_0, \ldots , y_{n-1}) \in \R^n: 0 \le y_i \le 1\},
    $$
     $$
     dI^n :=\{
(y_0, \ldots , y_{n-1}) \in \R^n: y_i = 0\text{ or }1\text{ for some }i\},
$$
     $$C((X,A),(Y,B)) = \{ f \in C(X,Y): f(A) \subset B\}.$$
Then $\pi_q(M) = [(I^q,dI^q),(M,m_0)]$.  Recall the 1--1 correspondence
between the sets of maps
$$\phi: C((I^n,dI^n),(\Omega_cM,m_0))\to C((I^{n+1},dI^{n+1}),(M,m_0)).$$
$$      \phi (f)(y_0,y_1 \ldots , y_n) = f(y_1, \ldots , y_n)(y_0)$$
(Here $m_0$ denotes both the base point of $M$ and the constant loop
onto it.)  It is well known that $\phi $ descends to an isomorphism on
the homotopy groups
  $$  \phi_*: \pi_n(\Omega_cM) \isomorphism \pi_{n+1}(M) .$$
Observe also that if $g \in
C((I^{q+1},dI^{q+1}),(M,m_0)) $ is differentiable then
 $$\phi^{-1}(g) \in
C((I^q,dI^q),(\Omega _dM,m_0)).$$
 So now we can show that
$$          (j\composition i)_*:\pi_q(\Omega _dM) \to \pi _{q}(\Omega_c M) $$
is bijective.  From 17.8 and 17.8.1 of Bott and Tu, \cite{BotTu}
 it follows that there is a
differentiable map, $g$, in the homotopy class of $\phi (f)$ (surjectivity
of $(j\composition i)_*$) and that any two differentiable
 maps, $\phi (f_0)$ and $\phi (f_1)$
which are continuously homotopic are homotopic via a path of
differentiable maps (injectivity of $(j\composition i)_*$).
 This argument also shows that
$j$ is a weak homotopy equivalence and thus so too is $i$.
\end{proof}

\subsection{The loop map}
If $X$ and $Y$ are two spaces (manifolds) and $f$ is a continuous
(differentiable) map
$               f: X \to Y $
then there is a continuous (differentiable) map, the loop of $f$, denoted
$               Lf: LX \to LY$ where
$                 \gamma \mapsto  f\composition \gamma$.
If $P(M,G,f)$ is a locally trivial principal $G$-bundle then
$LP(LM,LG)$ is a locally trivial principal $LG$-bundle.
Now, we may realise $EG(BG,G)$ as a smooth principal $G$-bundle via
the inclusion of $G$ in $O(n)$ for some $n$ and the realisation of the
classifying space of $O(n)$ as the infinite dimensional Steifel manifold
(see \cite{Whi}).  It follows that $LEG$ makes sense for differentiable loops
and since $LEG$ is also a contractible space that

$$               BLG = LBG.$$
Since the homotopy class of a continuous map between
manifolds always contains a differentiable map we may take the classifying
map of any principal $G$-bundle to be differentiable and hence $LP(LM,LG)$
has classifying map $Lf$.
All of this holds mutatis mutandis for the based loops.

\subsection{Transgression}
Given two spaces, $X$ and $Y$,
the slant product (see \cite{Spa} p. 287) is the product in general
(co)homology theories which corresponds to integration over the fibre of
$X \times Y$ in de Rham theory.
Let $\omega \in H^q(X \times Y, \Z)$, $a \in H_p(X,\Z)$
and $b \in H_{q-p}(Y,\Z)$ then the slant product:
$$ /:H^q(X \times Y,\Z) \times H_p(Y,\Z) \to H^{q-p}(X,\Z)$$
is given by
$$ (\omega / a)(b) = \omega(a \otimes b)$$
We shall need the following functorial property.  Given $f: X \to X'$,
$g: Y \to Y'$ and $\omega ' \in H^q(X' \times Y',\Z)$ then
\begin{equation}
\label{slantfunct}
[(f \times g)^*\omega ']/ a = f^*(\omega '/g_* a)
\end{equation}
Let $ev: \Omega X \times S^1 \to X$ be the evaluation map and let
$i$ be the fundamental class of $H_1(S^1,\Z)$.  Then
the transgression homomorphism between the cohomologies of a space and
its loop space is defined as follows.
$$
\begin{array}{ccc}
t^q: H^{q+1}(X, \Z) & \rightarrow & H^q(\Omega X, \Z) \\
\omega & \mapsto & ev^*(\omega)/i
\end{array}
$$
One can easily check that the following diagram commutes.
$$
\begin{array}{ccccc}
\Omega X \times S^1 & \stackrel{ev} {\rightarrow} & X & \stackrel
{f}{\rightarrow} & X'\\
& \stackrel{\Omega f \times Id}{\searrow}  & &  \stackrel{ev}{\nearrow} &\\
& & \Omega X' \times S^1& &
\end{array}
$$
By applying \ref{slantfunct} to $\Omega f \times Id$ and $Id$
one sees that
\begin{equation}
\label{trfunc}
\begin{split}
t^q(f^* \omega) &= (ev^*(f^* \omega))/i \\
                &= (\Omega f \times Id)^*(ev^*)(\omega)/i \\
                &= (\Omega f)^*((ev^*)(\omega)/Id_* i)\\
		&= (\Omega f)^*t^q(\omega).
\end{split}
\end{equation}

In simple cases, McLaughlin (\cite{Mcl} p 147) has noted that transgressions
can be computed using the Hurewicz homomorphism as follows.
Given  any spaces $X$ and $Y$, let $[X, Y]_0$ denote
the set of based homotopy classes of continuous based maps from $X$ to $Y$,
then  there is a well known bijective,
adjoint correspondence (closely related to the correspondence mentioned in
Proposition \ref{3.14prop})
\begin{equation}
\label{adjoint}
[\Sigma X,Y]_0 \stackrel{\Delta}{\longrightarrow} [X,\Omega_c Y]_0
\end{equation}
which descends in the case that $X = S^{q-1}$ to the isomorphism between
the homotopy groups of a space and its loop space,
$$ \delta_q: \pi_q(Y) \cong \pi_{q-1}(\Omega_c Y).$$
In fact, $\delta_q = \partial_q$, the boundary map in
the long exact sequence of the continuous path fibration, $P_cY \to Y$.
Now let $\pi: S^{q-1} \times S^1 \to \Sigma S^{q-1}$ be the projection
defined by the equivalence relation $(\theta,1) \sim (\theta' ,1)$
and $(\theta_0,t) \sim (\theta_0,t')$
for all $\theta ,\theta' \in S^{q-1}$ and for all $t,t' \in S^1$
where $\theta_0$ is the base point of $S^{q-1}$.
Then, by the definition of $\Delta$,
the following diagram commutes.
$$
\begin{array}{ccccc}
S^{q-1} \times S^1 & \stackrel{\Delta f \times Id}{\longrightarrow} &
\Omega_c X \times S^1 & \stackrel{ev}{\longrightarrow} & X \\
& \stackrel{\pi}{\searrow} & & \stackrel{f}{\nearrow} & \\
& & \Sigma S^{q-1} & &
\end{array}
$$
If $j \in H_{q-1}(S^{q-1},\Z)$ is a generator then
$\pi_*(j \otimes i) := k$ generates $H_q(S^q,\Z)$.  Thus
for $\omega \in H^q(X,\Z)$,

\begin{equation}
\label{trhure}
\begin{split}
\omega(f_*(k)) &= \omega(f_* \pi_*(j \otimes i)) \\
&= \omega(ev_*((\Delta f)_* j \otimes i)) \\
&= ev^*(\omega)((\Delta f)_* j \otimes i)\\
&= t^q(\omega)((\Delta f)_* j)
\end{split}
\end{equation}

In cases where the Hurewicz homomorphism, $\phi: \pi_{q-1}(\Omega_c X)
\to H_{q-1}(\Omega_c X,\Z)$ is surjective and $H^{q-1}(\Omega_c X,\Z)$
is torsion free, (\ref{trhure}) will allow us to compute $t^q$ since
in this case a cohomology class $\omega ' \in H^{q-1}(\Omega_c X, \Z)$ is
determined by the value it takes on $(\Delta f)_* j$ as $\Delta f$ runs
through $\pi_{q-1}(\Omega_cX)$.
We can also use the fact that the continuous and differentiable loop spaces
are homotopic (Proposition \ref{3.14prop}) to gain the same
result when $X= M$ is a manifold and we consider
differentiable loops (now we must consider a differentiable map, $g:S^{q-1}
\to \Omega_d M$ which is homotopic to $\Delta f$).

We can apply this to interpret the transgression homomorphism as the
looping of maps when we regard $H^q(X,\Z)$ as $[X,K(\Z,q)]$.
In this case the Hurewicz homomorphism is an isomorphism and if
$1 \in H^q(K(\Z,q),\Z)$ is a generator then $\tau^q(1) := 1'$ generates
$H^{q-1}(\Omega K(\Z,q),\Z) = H^{q-1}(K(\Z,q-1),\Z)$ and
\begin{equation}
\label{trloop}
\tau^q(f^*(1)) = (\Omega f)^*(1').
\end{equation}

\section{Killingback's result}
In this section we confine our attention to cases where $G$
is a compact, connected and simply connected  Lie group
and we consider string structures for smooth
bundles with fibre $\Omega_sG$.   We can consider $\Omega_s G$
and $\Omega_dG$
interchangeably since the obvious inclusion
$$\Omega_dG \inclusion \Omega_sG$$
is a homotopy equivalence.  This means that, for a Lie group $G$,
isomorphism classes of $\Omega_dG$-bundles, $\Omega_sG$-bundles
and $\Omega_cG$-bundles bundles are in 1--1
correspondence via the obvious bundle inclusions.
Thus, the problem of finding a string structure is identical in the case of
$\Omega_dG$ and $\Omega_sG$ as the following commutative diagram makes clear.
$$
\begin{CD}
        H^1(M,\Omega_sG) @.\quad \cong \quad @.H^1(M,\Omega_dG)\\
            @ V{D}VV        @.            @ V{D}VV\\
       H^2(M,\underline{S^1} )  @.\quad      \cong \quad
        @. H^2(M,\underline{S^1} )
\end{CD}
$$
We see that for a principal $G$-bundle, $P(M,G)$, over a manifold, $M$,
$D[\Omega_s P]=0$ if and only if $D[\Omega_dP]=0$.
This links the work of Carey and Murray \cite{CarMur} and McLaughlin
\cite{Mcl}.
Moreover since $LG$ is homeomorphic to $\Omega G \times G$ we need only
consider based loops when $G$ is simply connected for then
$H^i(G,Z) = 0$ for $i=1,2$ and the canonical projection
$\phi: LG = \Omega G \times G \to \Omega G$, induces an isomorphism
$$
\phi^*: H^2(\Omega G, \Z) \cong H^2(LG, \Z).
$$
The correspondence between circle bundles and second integral
cohomology entails,
$$
\widehat{L_s}(L_sG,S^1)=\phi^* \widehat{\Omega_sG}
(\Omega_s G, S^1) = \widehat{\Omega_sG}(\Omega_sG, S^1)
\times G.
$$
Now note that for any topological groups, $G$ and $H$, $H^1(M,\underline
{G \times H}) = H^1(M,\underline{G}) \times H^1(M,\underline{G})$.  The
following commutative diagram shows that
$L_sP$ has a string structure if and only if $\Omega_sP$ has one
(the first two vertical arrows are the obvious
projections and $\rho$ is the map induced on cohomology from the
projection $\rho: \widehat{\Omega_sG} \to \Omega_s G$).

$$
\begin{CD}
H^1(M,\underline{\widehat{\Omega_sG}}) \times H^1(M,\underline{G})
@>{\rho \times Id}>>
H^1(M,\underline{\Omega_sG}) \times H^1(M,\underline{G}) @>{D}>>
H^2(M,\underline{S^1}) \\
@VVV @VVV @V{Id}VV \\
H^1(M,\underline{\widehat{\Omega_sG}}) @>{\rho}>>
H^1(M,\underline{\Omega_sG}) @>{D}>> H^2(M,\underline{S^1})
\end{CD}
$$

Let us now turn to the general situation for $\Omega_sG$.
Start with a principal $SO(n)$--bundle, $P(M,SO(n),f)$ ($n > 4)$,
(typically $P$ is the frame bundle of the tangent bundle of a
Spin manifold $M$) that has a
$Spin(n)$--structure $Q(M,Spin(n),\Hat{f})$ and form the loop bundle
$\Omega_sQ(\Omega_sM,\Omega_s Spin(n),\Omega_s \hat{f})$.  Now, realise
$B\Omega_s Spin(n)$ as $\Omega_sBSpin(n)$.  Since $Spin(n)$ is two-connected
with $\pi_3(Spin(n)) \cong \Z$, $BSpin(n)$ is
three-connected and $H^4(BSpin(n),\Z) \cong \Z$.
Thus (\ref{trhure}) gives us that
$$t^4: H^4(BSpin(n),\Z) \to H^3(\Omega_s BSpin(n),\Z)$$ is
an isomorphism so choose $\omega \in H^4(BSpin(n),\Z)$, a generator, so that
$t^4(\omega) = \mu$, the universal Dixmier-Douady class.  So,
\begin{align*}
D[L_sQ] &= (\Omega_s \hat{f})^* \mu \\
&= (\Omega_s \hat{f})^* t^4(\omega) \\
&= t^4(\hat{f}^*\omega) \text{             by (\ref{trfunc})}.
\end{align*}
McLaughlin \cite{Mcl} in his Lemma 2.2 shows by analysing the spectral
sequence of the bundle $BSpin(n)(BSO(n),B\Z_2)$ that for $n>4$
$$        2.\hat{f}^*(\omega) = P_1(P),$$
where $P_1(P)$ is the first Pontryagin class of $P$.  Thus
$$ 2D[L_sQ] = t^4(P_1(P))$$
which is Killingback's result.  Now (\ref{trhure}) entails that $t^q$
is injective for $M$ $(q-2)$-connected and hence the vanishing of
$(1/2)P_1(P)$ is a necessary and sufficient for the existence of a string
structure if $M$ is two-connected, and merely a sufficient condition in
general.

\section{The restricted unitary group}
\label{res-unitaries}
We start with a separable Hilbert space ${H}
= {H}^+ + {H}_-$ decomposed
by infinite dimensional subspaces
${H}^+$ and ${H}_-$ which are
the range of the self adjoint projections $P^+$ and $P^-$
respectively,
$Id_{H}  = P^+ + P^-$.  The restricted unitary group
relative to a
polarisation is defined by
$$      \Ur ({H} ,P^+) = \{ u \in U({H} ):
P^\pm uP^\mp \text{ is Hilbert Schmidt} \}. $$
Now $\Ur$  is not equipped with the subspace topology from
$U({H} )$ but with its own topology coming from the metric
$\rho $.
\begin{align*}
\rho (u_1,u_2) &= ||P^+(u_1-u_2)P^+|| + ||P^-(u_1-u_2)P^- || \\
                         &+ |P^+(u_1-u_2)P^-|_{HS} +
|P^-(u_1-u_2)P^+|_{HS}
\end{align*}
Where $|\quad |_{HS}$ denotes the symmetric norm on the
Hilbert--Schmidts.
Typically the Hilbert space and polarisation are understood and
omitted from the notation.  If $(\ ,\ )$ denotes the inner product on
${H}$, then the $CAR$ (canonical anti-commutation relations)
algebra over ${H}$, $CAR({H} )$ is
the $C^*$--algebra
generated by the set
$$               \{ a(f),a^*(f),f \in {H} \} $$
whose elements satisfy the canonical anti--commutation relations
$$               a(f).a(g) + a(g)a(f) = 0 $$
$$               a(f).a^*(g) + a(g^*).a(f) = (f,g).$$
  Any unitary $u \in U({H} )$ allows one to define an
automorphism of $CAR({H} )$ (called a Bogoliobuv transformation)
by
$$         \alpha_u((a(f)) = a(u.f) \quad \alpha_u((a^*(f)) =
a^*(u.f) $$
An irreducible  (Fock)
 representation $\pi$ of $CAR({H} )$ is determined via the GNS
construction from
the state $\omega$ defined by
$$\omega (a^*(f_1)...a^*f_M)a(g_N)...a(g_1)=\delta_{MN}
det(g_i,P^-f_j).$$
 The result we need, due originally to Friedrichs, is the theorem
(see \cite{ShaSti})
 that, given a Bogoliubov
transformation $\alpha_u$, there exists a unitary $W(u) \in
U({H}_\pi )$ such
that
$$ \pi (\alpha_u(a(f)))=\pi (a(u.f))=Ad(W(u))(\pi (a(f))=W(u)\pi
(a(f))W(u)^*$$
iff $u \in  \Ur ({H} )$.  Since $\pi$ is irreducible,
$W(u)$, is uniquely
defined up to a scalar which is killed by the adjoint.  Hence the
above defines an embedding
$$               i:  \Ur  \inclusion \PU({H}_\pi ) $$
of the restricted unitaries of ${H}$  in the projective
unitaries on
${H}_\pi $.  It is a corollary of a proof of (Carey 1984
Lemma 2.10) that this embedding is closed in $\PU({H}_\pi
)$.  Furthermore
we shall see below that $H^2( \Ur ,\integers) = \integers$ and
the canonical central
extension of  $\Ur $, $ \widehat{U}_{res}$, defined by the generator
of
$H^2( \Ur ,\integers)$ is given by
$$ \Ur ( \Ur ,S^1) = i^*U({H}_\pi )(\PU({\mathcal
H}_\pi ),S^1).$$
Hence the assumptions of Section 3 are fulfilled.
Finally, note that $\Ur$ is a disconnected group with
connected components labelled by the Fredholm index of $P^+UP^+$.
We denote the
connected component of the identity by $\Ur^0 $.
Henceforth we
drop reference to the different Hilbert
spaces over which $ \Ur$  and $\PU$ are defined and it shall be
understood that $\PU$ refers to the projective unitaries on ${\mathcal
H}_\pi$  and
not ${H} $.

We now summarise the
homotopy properties of  $\Ur $ and its role as a classifying
space for
$U(\infty )$ and the relation between  $\Ur$  and $\PU$ bundles.

\section{$\Ur $ as a classifying space}
\label{3.16section}
The group of unitaries with determinant, say ${\mathcal T} $,
consists of those operators of the form $1+$trace class.
 By considering ${\mathcal T} ({H} +)$,
Pressley and Segal \cite{PreSeg} (see
Ch 6) show
that there is a principal ${\mathcal T}$--bundle over $ \Ur^0 $,
the connected component of $\Ur$
with contractible total
space and hence $ \Ur^0 $ is a $B{\mathcal T} $.
It is known that ${\mathcal T} $ has
homotopy type of the direct limit of the finite unitaries.
$${\mathcal T}  \homotopyequivalence  U(\infty ) =
\lim_{n\to\infty }
(U(n)) $$
So  $\Ur^0$ is a $CW$--classifying space for
$\mathcal T$ and thus $U(\infty )$. So we have $\Ur^0
\homotopyequivalence B{\mathcal T} $.  Since the homotopy
groups of $U(\infty )$ are well known by Bott periodicity we have that
$$         \pi_q( \Ur ) = \begin{cases}\integers, & q\text{
even},\\ 0 &q \text{ odd}.\end{cases} $$
(This result has elsewhere been proven via methods more closely
tied to  $\Ur $'s structure as a group of operators, see Carey
(1983).)  Now $U(\infty )$ and $BU(\infty )$
 are classifying spaces for
reduced $K$--theory and we have:

\begin{proposition}
\label{3.17prop}
$$B \Ur  \homotopyequivalence  U(\infty ) \quad
 \Ur \homotopyequivalence \Omega_c U(\infty ) .$$
\end{proposition}
\begin{proof}
It is known that the embedding of
$\Omega_dU(n) \inclusion \Ur$ extends to a map $i:\Omega _dU(\infty )
\inclusion \Ur$ and one can
check that this is a weak homotopy equivalence and hence a homotopy
equivalence.  By Proposition \ref{3.14prop},
$\Omega _dU(\infty ) \homotopyequivalence \Omega _cU(\infty ) $
and thus,
remembering that via the path fibration $B\Omega _cG \simeq G^0$,
$$      B\Ur \homotopyequivalence B\Omega _cU(\infty ) = U(\infty
).$$
If we loop this equation we find,
$$      \Omega _cU(\infty ) \homotopyequivalence \Omega _cB\Ur
\simeq B\Omega_c\Ur \simeq \Ur. $$
\end{proof}

\begin{note}
\label{note(a)} Over the category of CW-complexes of dimension less
than a given integer, $CW_n$, and over the category of finite CW-complexes,
$CW_{fin}$, the
functors of reduced K-theory have $BU(\infty)$ as a classifying space
(see \cite{Hus} p118).  If follows that isomorphism classes of
$\Ur$-bundles correspond bijectively with elements of reduced
$K$-theory.
Specifically, $\tilde{K}^1 (X)$ of a base is defined to be
the stable isomorphism classes of vector bundles over the
reduced suspension of $X$, $\Sigma  X$.  For $X \in CW_n$ or $CW_{fin}$

\begin{align*}
         \tilde{K}^1(X) &= [\Sigma X,BU(\infty )]\\
                &= [X,\Omega_{c}BU(\infty )] \text{        apply $\Delta$}\\
                &= [X,B\Omega_cU(\infty )]\\
                &= [X,B \Ur ]\\
                &= Bun_X( \Ur )
\end{align*}
where $Bun_X( \Ur )$ denotes the set
of all isomorphism classes of $\Ur$ bundles
over $X$. Now elements of $\tilde{K}$ correspond bijectively with
$U(\infty)$-bundles,  $\Ur$-bundles correspond with
$\Omega _cU(\infty )$ bundles. So our correspondence can be seen as
a mapping between $\Omega _cU(\infty )$-bundles over a space and
$U(\infty )$-bundles over the reduced suspension of that space which
is attained by applying $\Delta$ or $\Delta^{-1}$ to the classifying
maps of the bundles.  We exploit this in the next subsection.
\end{note}

\section{The Dixmier-Douady class and the second Chern class}
\label{3.18}
Regarding $\Ur$ as a subgroup of $\PU$ via the inclusion mentioned
in
Section \ref{3.13section}, we may ask
when can we reduce the structure group of a $\PU$--bundle,
$P(\PU,M,f)$ to $\Ur$?
By Theorem \ref{th:obstruction} we translate this question into
a search for
maps $\hat{f}$ such that $f=g \composition \hat{f}$.  Where we take
$g:B\Ur \to B\PU$ to be a fibration with fibre $F$.
$$
\begin{array}{ccc}
& & B\Ur  \homotopyequivalence U(\infty ) \\
& & g \downarrow\\
X & \stackrel{f}{\longrightarrow} &B\PU \homotopyequivalence  K(\integers,3)
\end{array}
$$

In general we know that if there were a section of $\pi$, say $s$, then
this
would entail the existence of group homomorphisms
$$    g^*: H^*(B\PU,\integers) \to H^*(B\Ur,\integers) $$
$$     s^*:  H^*(B\Ur,\integers) \to H^*(B\PU,\integers) $$
such that $$ s^*\composition g^* = (g\composition s)^* = id. $$
It is a group theoretic result that this implies that
$H^*(B\PU,\integers)$
would be a direct summand of $H^*(B\Ur,\integers)$.  But we know
(See Bott
and Tu pp~245--246) that $H^*(B\PU,\integers)$ has torsion whereas
$H^*(B\Ur,\integers)$ is a free group.  Therefore the sought
after section
cannot exist and the structure group of some $\PU$--bundles does not
reduce to $\Ur$.

     The situation in specific instances depends in part on the
homotopy groups of the fiber, which we can compute in this case by
noting that
$$
i_{*,q}:\pi_q(\Ur) \to \pi_q(\PU)
$$
is an isomorphism
for $q=2$ and null otherwise.  It follows by Lemma \ref{lma:HintoG} that
$g_{*,q}:\pi_q(B\Ur) \to \pi_q(B\PU)$ is an isomorphism for $q=3$
and null otherwise.  By considering the long exact homotopy sequence
of the fibration
$$ F\inclusion BH \overset{g} \to BG$$
we see that
$$         \pi_q(F) = \begin{cases}\Z, & q\text{ odd $\neq
3$},\\ 0 &q \text{ even or $3$}.\end{cases} $$
Now  the cohomology, $H^n(K(\Z,3))$, of $K(\Z,3)$ is zero for
$n = 1$ and torsion for $n > 3$  (see Bott and Tu pp 245--246).
Hence obstructions to lifting $f$ can
lie only in $H^{2n+4}(M,\Z) $ ($n \ge 1$).  So the structure group of any
 $\PU$--bundle over a space with free, even (greater than fourth) cohomology
groups reduces to $\Ur$.

We recast this problem
 in a more general setting by exploiting the correspondence
between $\Ur$-bundles over a space $x$ and $\tilde{K}^1(X)$.
There is a suspension isomorphism on cohomology,
$$
\Sigma ^q: H^q(X,\integers ) \isomorphism H^{q+1}(\Sigma X,\Z)
$$
which one can obtain from the Mayer-Vietoris sequence for
$(\Sigma X,CX,CX)$
(where ``$CX$'' denotes the reduced cone of $X$) or by
using the adjoint
relation, $\Delta$ (see \ref{adjoint})
between $\Sigma$  and $\Omega _c $ considered as functors on $CW$:
$$
X \overset{\Delta f}{\longto } \Omega _cK(\integers ,q+1) = K(\integers ,q)
\longleftrightarrow
\Sigma X \overset{f}{\longto }K(\integers ,q+1).
$$
If $1$ and $1'$ are as in (\ref{trloop}) then
\begin{equation}
\label{susp-loop}
\Sigma^q((\Delta f)^*(1')) = f^*(1).
\end{equation}
The next proposition uses the
suspension isomorphism and the transgression homomorphism
to link characteristic classes of principal $\Ur$--bundles
over with the characteristic classes of the associated
$U(\infty )$--bundles over $\Sigma X$.

\begin{proposition}
\label{trcharprop}
Let $c$ be a characteristic class for principal $U(\infty)$-bundles
defined by its universal class $c^* \in H^{q+1}(BU(\infty),\Z)$ and let
$t^q(c)$ be the characteristic class for principal
$\Ur$-bundles with universal class $t^q(c^*)$.
If $P(X,\Ur,\Delta f)$ is a principal $\Ur$--bundle over
$X$ and $$\Sigma P(\Sigma X,U(\infty), f) )$$
is the associated $U(\infty )$--bundle over $\Sigma X$,
>(element of $\tilde{K}^1(X)$)then
$$      c(\Sigma P) = \Sigma^q(t^q(c)(P)).$$
\end{proposition}
\begin{proof}

Let $c$ also denote a map $c:BU(\infty) \to K(\Z,q+1)$ which pulls back
$1 \in K(\Z,q+1)$ to $c^*$.  We must show that $f^*(c^*) =
\Sigma^q((\Delta f)^*(t^q(c^*)))$.
By applying Proposition \ref{3.17prop}, realise
$B\Ur$ as $\Omega_c BU(\infty)$.  Thus we may exploit the
adjoint pairing $\Delta$ between the functors $\Omega_c$ and $\Sigma$.
We have the following maps
$$
\begin{array}{ccccc}
\Sigma X & \stackrel{f}{\longrightarrow} & BU(\infty)
& \stackrel{c}{\longrightarrow} & K(\Z,q+1) \\
X & \stackrel{\Delta f}{\longrightarrow} & \Omega_c BU(\infty)
& \stackrel{\Omega_c c}{\longrightarrow} & \Omega_c K(\Z,q+1)
\end{array}
$$
with $\Delta (c \circ f) = \Omega_c c \circ \Delta f$.  Now by
\ref{trloop}, $(\Omega_c c)^*(1') = t^q(c^*)$ and thus
\begin{align*}
\Sigma^q((\Delta f)^*(t^q(c^*)))
&= \Sigma^q((\Delta f)^*(\Omega_c c)^*(1')) \\
&= \Sigma^q(\Delta (c \circ f)^*(1')) \\
&= (c \circ f)^*(1) \text{      by \ref{susp-loop}} \\
&= f^*(c^*)
\end{align*}
and the proposition is proved.
\end{proof}

\begin{proposition}
Let $D$ be the Dixmier-Douady class for principal $\Ur$-bundles,
let $c_2$ be the second Chern class for $U(\infty)$-bundles and let
$P$ and $\Sigma P$ be as above.  Then
$$ \Sigma^3(D(P))= c_2(\Sigma P)$$
\end{proposition}
\begin{proof}
Let $D^* \in H^3(B\Ur,\Z)$ and $c_2^* \in H^4(BU(\infty,\Z))$ denote
the universal classes of $D$ and $c_2$ respectively, then by Proposition
\ref{trcharprop} it suffices to show that
$$ t^3(c_2^*) = D^*. $$
Using \cite{Hus} (Ch 20, Corollary 9.8) one deduces that there is there is a
$U(\infty))$-bundle over $S^4$,
$P(S^4,U(\infty),f)$, with $c_2(P)$ a generator of $H^4(S^4,\Z)$.
Let $k$ denote the generator of $H_4(S^4,\Z)$ such that $1 = c_2(P)(k)
= f^*(c_2^*)(k)$ and let $j$ be the corresponding generator of
$H_3(S^3,\Z)$ (in the sense of \ref{trhure}).  Then by (\ref{trhure})
$$t^3(c_2^*)((\Delta f)_*j) = f^*(c_2^*)(k) = 1.$$
But one can show by considering long exact sequence of the fibration
$$U(n)(S^{2n-1},U(n-1))$$ ($n$ large) that the Hurewicz homomorphism is an
isomorphism on
$$
\pi_3(B\Ur) \cong \pi_3(U(\infty )) \cong \Z.
$$
Hence, $t^3(c_2^*)$
generates $H^3(B\Ur,\Z)$ (as it evaluates to $1$ on the generator of
$H_3(B\Ur,\Z)$) and so $t^3(c_2^*) = D^*$ as required.
\end{proof}

In summary, the structure  group of a $\PU$-bundle, $Q(X,\PU)$ reduces to
$\Ur$ if and only if there is a $\Ur$ bundle, $P(X,\Ur)$ whose
Dixmier-Douady
class coincides with that of $Q$.  This, we have just seen, happens if and
only if there is a
$U(\infty )$--bundle, $\Sigma P(\Sigma X,U(\infty ))$ over
$\Sigma X$ such that $c_2(\Sigma P)
= \Sigma^3 (D(P))$.  We know from above that one cannot, in general,
construct a $U(\infty )$-bundle with an arbitrary second Chern class
on any given space.  This differs from the case for the first Chern class
where one can always find a line bundle, and hence a $U(\infty)$--bundle,
for any given element of $H^2(M,\Z )$.

\section{Connections with other viewpoints}

\subsection{Bundle gerbes}

An alternative method of defining the obstruction
to lifting a bundle to a central extension is to
use the notion of {\em bundle gerbes} \cite{Mur}.
We will sketch the theory here and refer the
reader to \cite{Mur} for details.
If $Y \to M$ is a fibration define $Y^{[p]}$ to
be the $p$th fibre product of $Y$ with itself.
Then a bundle gerbe over $M$ is a pair $(J, Y)$
where $\pi \colon Y \to M$ is a fibration and
$J \to Y^{[2]}$ is a $U(1)$ bundle.
Furthermore  for any $x$, $y$ and $z$ in $Y$
we require the existence of a bundle morphism, called the
bundle gerbe product,
$$
J_{(x, y)} \otimes J_{(y, z)} \to J_{(x, z)}
$$
depending continuously or smoothly on $x$, $y$ and $z$.
Moreover this composition is required to be associative.
Note that for $U(1)$ principal bundles there is a
natural notion of tensor product and dual, see \cite{Mur} for details.

If $L \to Y$ is a $U(1)$ bundle then we can define a
bundle gerbe $(Y, \delta(L))$  by
$$
\delta(Y)_{(x, y)} = L_x \otimes L_y^*.
$$
A bundle gerbe is called
trivial if it is isomorphic to a a bundle
gerbe of the form $\delta(L)$. The obstruction
to a bundle gerbe $(J, Y)$ over $M$ being {\em trivial}  is a
three class in $H^3(M, \Z)$ called the Dixmier-Douady
class of the bundle gerbe.  Its definition
can be found in \cite{Mur}.

The connection with our work is the
 bundle gerbe arising
as the obstruction to extending the
structure group of a $G$ bundle $P$ to $
\hat G$ where
$$
0 \to U(1) \to \hat G \to G \to 0
$$
is a central extension.
Note that if we form the fibre product $P^{[2]}$
there is a map $\sigma \colon P^{[2]} \to G$
defined by $p = q \sigma(p, q)$. We define
$J = \sigma^*(\hat G)$ where here we think  of $\hat G$
as a $U(1)$ bundle over $G$. It is easy to check
that the group multiplication in $\hat G$ defines the
required bundle gerbe product. It is shown in \cite{Mur}
that
\begin{theorem}[\cite{Mur}]
The bundle gerbe $L$ is trivial if and only if
the bundle $P$ lifts to $\hat G$.
The Dixmier-Douady class of  $L$
is the same Dixmier-Douady class which is the obstruction
to the bundle $P$ lifting to $\hat G$.
\end{theorem}

\subsection{The Dixmier-Douady class and Clifford bundles}

We now interpret the Dixmier-Douady class as an obstruction in
a different setting which is closer in spirit to that of the
original (cf \cite{Dix}).  Suppose we have a principal fibre bundle
$P(M,\Ur)$
and a locally finite cover $\{ U_\beta | \beta \in   A \} $ of $M$.
The transition
functions
$ g_{\beta\gamma } $
may be used to define the transition functions for a locally trivial
bundle
over $M$ with fibre the CAR algebra.
This is achieved by defining automorphisms of the CAR algebra by
$u_{\beta\gamma }(a(v))=a( g_{\beta\gamma }v)$ ($v \in {\cal H}$)
and using the $u_{\alpha\beta }$ as transition functions
for a fibre bundle $C(M,CAR({\cal H}))$.
If the Dixmier-Douady class of $P(M,\Ur)$ is trivial
then one can find unitaries
$$
\{W(u_{\beta\gamma })|\ \beta, \gamma \in A \}
$$
acting on the Hilbert space ${H}_\pi$
of $\pi$ which form a Cech 2-cocycle with values
in the unitaries on ${H}_\pi$.
Using these as transition functions one defines a `Fock bundle'
over $M$. with fibre the Fock space ${H}_\pi$.
Thus the Dixmier-Douady class of
$P(M,\Ur)$ is an obstruction to the existence of a locally trivial bundle
over $M$ with fibre the Fock space and on sections of which the Clifford bundle
(as a field of C$^*$-algebras) acts. This is analogous to the original
introduction of the Dixmier-Douady class as
an obstruction to the triviality of a bundle of C$^*$-algebras
with fibre the compact operators.

\bigskip

\noindent{\small \bf ACKNOWLEDGMENTS} \\ ALC and MKM
acknowledge  the support of the Australian Research Council.
We  thank Dr Mathai Varghese and  Dr Jim Davis for
assistance. Finally we thank John Phillips for suggesting many
improvements to the arguments.

1. Department of Pure Mathematics, University of Adelaide,
Adelaide, South Australia 5005, Australia.
{\tt acarey@maths.adelaide.edu.au},
{\tt mmurray@maths.adelaide.edu.au}

2. Department of Mathematics,
Indiana University,
Bloomington, IN 47405-4301,
USA. {\tt dcrowley@iu-math.math.indiana.edu}

\end{document}